\documentclass[reprint, showpacs,amsmath,amssymb, aps,prl]{revtex4-1}

\usepackage{graphicx}
\usepackage{dcolumn}
\usepackage{bm}

\newcommand{\ryd}{$\mathrm{5}s\mathrm{56}d\,^\mathrm{1}D_\mathrm{2}\ $}
\newcommand{\dt}{$\Delta t\ $}
\newcommand{\intstate}{$\mathrm{5}s\mathrm{5}p\,^{\mathrm{1}}P_{\mathrm{1}}\ $}
\newcommand{\fstate}{$\mathrm{5}s\mathrm{54}f\,^{\mathrm{1}}F_{\mathrm{3}}\ $}
\newcommand{\pstate}{$\mathrm{5}s\mathrm{56}p\,^{\mathrm{1}}P_{\mathrm{1}}\ $}
\begin{document}


\title{Two-electron excitation of an interacting cold Rydberg gas}

\author{J. Millen}
\author{G. Lochead}
\author{M. P. A. Jones}
\email{m.p.a.jones@durham.ac.uk}
\affiliation{Department of Physics, Durham University, Durham DH1 3LE, United Kingdom}

\date{\today}

\begin{abstract}
We report the creation of an interacting cold Rydberg gas of strontium atoms. We show that the excitation spectrum of the inner valence electron is sensitive to the interactions in the Rydberg gas, even though they are mediated by the outer Rydberg electron. By studying the evolution of this spectrum we observe density-dependent population transfer to a state of higher angular momentum $l$. We determine the fraction of Rydberg atoms transferred, and identify the dominant transfer mechanism to be $l$-changing electron-Rydberg collisions associated with the formation of a cold plasma.  
\end{abstract}

                           
\maketitle

The combination of laser-cooled atoms and the strong, tunable, long-range dipole-dipole interactions between Rydberg states has emerged as a powerful new way of studying strongly interacting quantum systems. In particular, the laser excitation of Rydberg states in cold, dense clouds can lead directly to the formation of highly entangled states via the dipole blockade mechanism \cite{lukin01, saffman09, gaetan09, heidemann07}. The interatomic interactions also modify the atom-light interaction \cite{schempp10}, giving rise to cooperative effects \cite{pritchard10} that could be exploited to create photonic quantum gates \cite{friedler05}. Novel binding mechanisms give rise to long-range molecules \cite{bendkowsky09, overstreet09}, and the Rydberg gas has also been observed to spontaneously evolve into a plasma \cite{robinson00}.

Currently, these experiments use alkali metal atoms, where only the Rydberg electron is available for manipulation by external fields. Elements that have two valence electrons, such as strontium, provide additional degrees of freedom. The valence electron that remains in the core can be excited independently \cite{cooke78}. This raises the possibility of imaging \cite{simien04} or laser cooling Rydberg atoms, or confining them in an optical dipole trap. A cold gas of ``planetary'' Rydberg atoms \cite{percival77} could be created, where electronic correlations play an important role. The creation of highly entangled states of two-electron atoms via the Rydberg blockade could have important applications in precision frequency metrology \cite{blatt08}.

In this paper we report the creation of a cold Rydberg gas of strontium atoms. By exciting the inner valence electron, we probe the state of the outer Rydberg electron through the overlap of the two electronic wavefunctions. This overlap causes the doubly excited Rydberg atoms to autoionize. By measuring how the shape of the autoionizing resonance varies with density and time, we observe the transfer of population from the initially excited Rydberg state into other angular momentum states, in the presence of repulsive van der Waals interactions.  Several population transfer mechanisms have been identified in cold Rydberg gases, including black-body radiation \cite{beterov09}, superradiance  \cite{day08}, resonant energy transfer due to dipole-dipole interactions \cite{anderson98}, and electron-Rydberg collisions \cite{dutta01}. From a quantitative analysis of the spectra we determine that angular momentum ($l$) changing collisions associated with the formation of an ultra-cold plasma are the dominant mechanism for population transfer under our experimental conditions.
 
\begin{figure}
\includegraphics{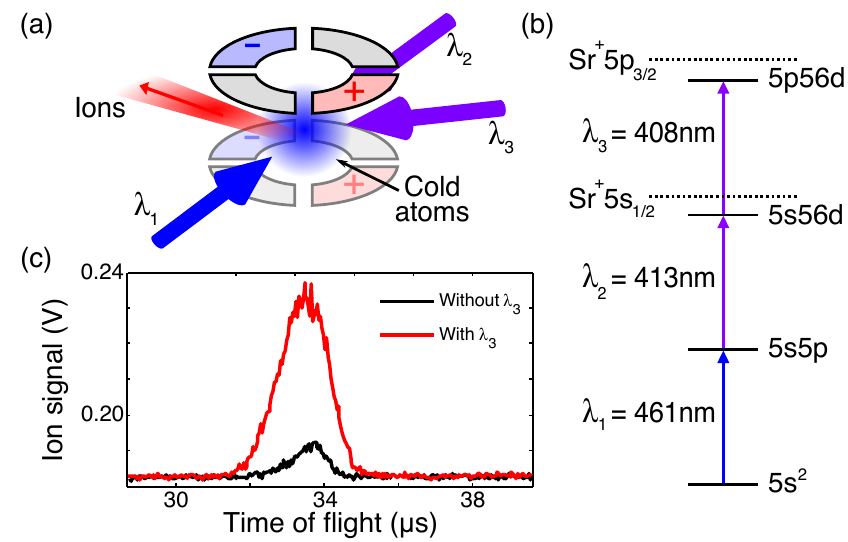}
\caption{\label{fig:fig1} (Color online) (a)~Schematic of the experiment. The MOT is formed between two segmented ring electrodes, which direct the ions towards the  MCP. (b)~Energy level diagram. The first electron is excited to the \ryd state by a two-step excitation ($\lambda_{\mathrm{1}} + \lambda_{\mathrm{2}}$). After a variable delay \dt, the second electron is excited at $\lambda_{\mathrm{3}}$ and the atom autoionizes. (c)~An example of the time-resolved ion signal. The autoionization signal is much larger that the signal from spontaneously created background ions.}
\end{figure}
 
A schematic of the experiment is shown in Fig.~\ref{fig:fig1}(a). The cold Rydberg gas is created by exciting cold atoms from a magneto-optical trap (MOT), which operates on the $\mathrm{5}s^\mathrm{2}\,^{\mathrm{1}}S_{\mathrm{0}}\to\mathrm{5}s\mathrm{5}p\,^{\mathrm{1}}P_{\mathrm{1}}$ transition at $\lambda_{\mathrm{1}}=\mathrm{461}$~nm. The MOT is loaded using a Zeeman slowed atomic beam, and we obtain  $\sim\mathrm{3}\times10^\mathrm{6}$ atoms at a temperature of 5~mK and a density of $\sim\mathrm{2}\times\mathrm{10}^{\mathrm{10}}$~cm$^{\mathrm{-3}}$. Once the MOT is loaded, the MOT and Zeeman slower light is extinguished, and the atoms are excited to the \ryd Rydberg state using the two-step excitation shown in Fig.~\ref{fig:fig1}(b). The first step is provided by a resonant probe beam  derived from the cooling laser, which is stabilized using polarization spectroscopy \cite{javaux10, bridge09}, and has an intensity of 0.7~I$_\mathrm{sat}$. The second step at $\lambda_\mathrm{2}$ = 413~nm is driven by a CW frequency-doubled diode laser system. These laser beams are counter-propagating, have a waist of 0.7~mm, and are linearly polarized in the vertical direction. Both lasers are pulsed on simultaneously for a duration of 4~$\mu$s. 

After a variable delay $\Delta t$, we excite the second electron, using a 4~$\mu$s pulse with a variable detuning $\Delta_\mathrm{3}$ from the Sr$^+$ $\mathrm{5}s_{\mathrm{1/2}}\to \mathrm{5}p_{\mathrm{3/2}}$ transition at $\lambda_{\mathrm{3}} = \mathrm{408}$~nm. This light is provided by a CW external cavity diode laser, and the beam has a power of 1.3~mW and a waist of $\sim$~1~mm. This pulse autoionizes the Rydberg atoms, and is followed by a 4~$\mu$s electric field pulse that directs ions to a micro-channel plate (MCP) detector. The amplitude of the field pulse (3.6~V\,cm$^{\mathrm{-1}}$) is not sufficient to field ionize the Rydberg atoms. The ion signal is averaged over 100 cycles of the experiment. An example of the averaged, time-resolved ion signal is shown in Fig.~\ref{fig:fig1}(c), which illustrates the increase in signal due to autoionization. 

For each value of the detuning $\Delta_{\mathrm{3}}$ and delay \dt, we measure the spectrum of the Rydberg excitation by stepping the wavelength $\lambda_{\mathrm{2}}$ across the Rydberg transition. In this way, we obtain the peak ion signal on resonance $S$, which we normalize by the atom number and power of the 408~nm beam.  The linewidth of the Rydberg excitation is close to the 32~MHz width of the \intstate intermediate state, and we see no evidence of density-dependent broadening effects \cite{anderson98,tanner08,reetz08}.

\begin{figure}
\includegraphics{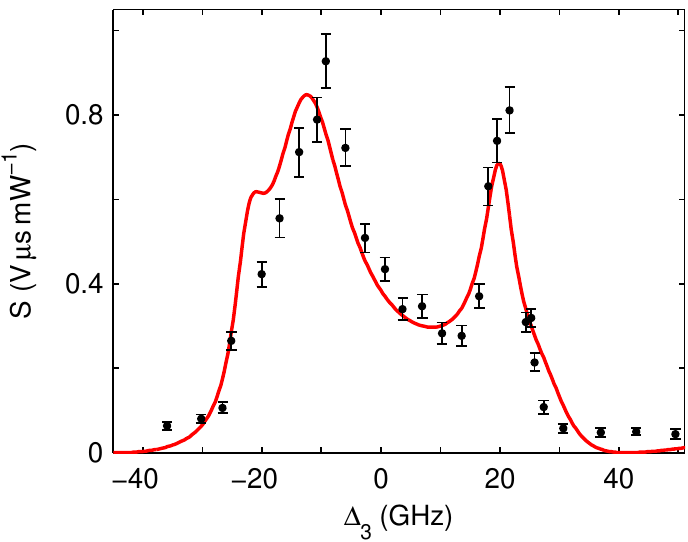}
\caption{\label{fig:fig2} (Color online) The autoionization signal at $\Delta t = \mathrm{0.5}\ \mu$s as a function of the detuning $\Delta_{\mathrm{3}}$, for $P_{\mathrm{2}} = \mathrm{1}$~mW. The solid line is a fit using a six-channel MQDT model for the $\mathrm{5}s\mathrm{56}d$ state.}
\end{figure}

\begin{figure}
\includegraphics{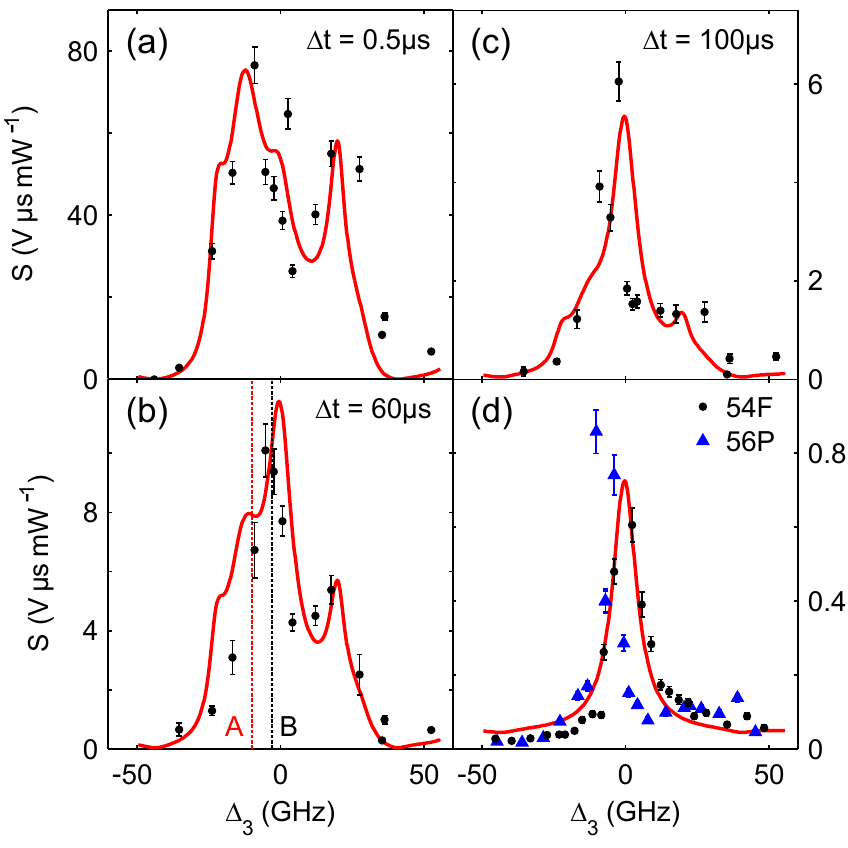}
\caption{\label{fig:fig3} (Color online). (a--c) Autoionization spectrum for three different values of $\Delta t$. The Rydberg laser power $P_{\mathrm{2}} =\mathrm{10}$~mW. The background ionization signal has been subtracted from the data. The solid lines are fits to the data using the combined $\mathrm{5}s\mathrm{56}d$ and $\mathrm{5}s\mathrm{54}f$  MQDT model. (d) Autoionization spectra for the \fstate (black dots) and \pstate (blue triangles) Rydberg states. The solid line is a two-channel MQDT fit to the $\mathrm{5}s\mathrm{54}f$ data.}
\end{figure}

However, density-dependent effects {\em are} visible in the excitation spectrum of the inner electron. Figure~\ref{fig:fig2} shows the autoionization spectrum at $\Delta t = \mathrm{0.5}\ \mu$s, taken with a low power ($P_{\mathrm{2}}=\mathrm{1}$~mW) 413~nm beam. The spectrum exhibits a double-peaked structure that is characteristic of the Sr $\mathrm{5}pnd$ autoionizing states \cite{cooke78}, and which can be well described by six-channel multi-channel quantum defect theory (MQDT) \cite{xu87}.  This model is in good agreement with with our data in Fig.~\ref{fig:fig2}. As the power $P_{\mathrm{2}}$, and thus the Rydberg density, is increased, the lineshape changes, as shown in Fig.~\ref{fig:fig3}(a). By increasing the delay $\Delta t$, the emergence of an additional feature becomes clear. As shown in Figs.~\ref{fig:fig3}(b-c), the shape of the Rydberg spectrum clearly evolves as $\Delta t$ is increased. The double peak structure associated with the $\mathrm{5}s\mathrm{56}d$ state decays, and after 100~$\mu$s only this narrow density-dependent feature remains. Therefore, as the density of Rydberg atoms is increased, we are observing the autoionization of Rydberg atoms in states other than the initially excited $\mathrm{5}s\mathrm{56}d$ state \cite{cooke79}.

The reduction in linewidth as compared to the $5s56d$ state indicates a reduction in the overlap between the two electrons. This could, in principle, result from an increase in principal quantum number $n$ or $l$.  Population transfer to states nearby in $n$ does not significantly alter the spectrum, as the width scales as $n^{-\mathrm{3}}$. The narrow peak that we observe would require the selective population of Rydberg states with $n\ge\mathrm{70}$. Higher $\mathrm{5}snd$ states would also give rise to a double-peaked structure, which we do not observe in Fig.~\ref{fig:fig3}. In contrast, the width of the autoionizing spectrum decreases extremely rapidly with $l$ \cite{jones90}, until for $l\ge\mathrm{8}$ the inner electron is more likely to decay by spontaneous emission.

\begin{figure}
\includegraphics{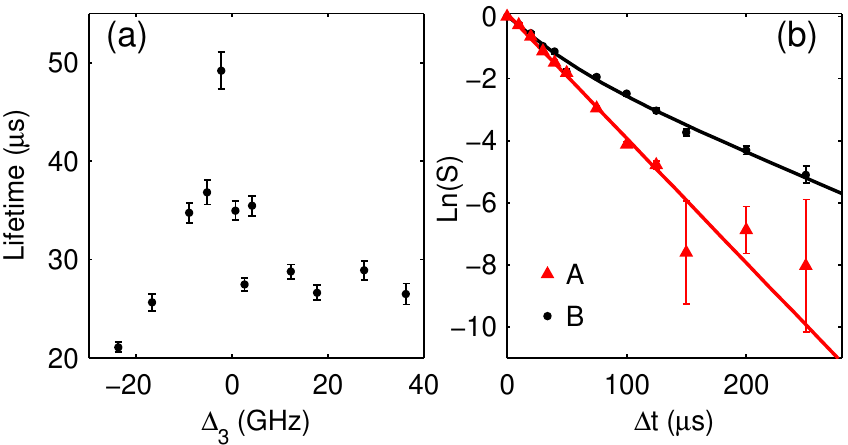}
\caption{\label{fig:fig4}(Color online) (a) Variation in the lifetime, derived from a single exponential fit to the decay of the autoionization signal, with $\Delta_{\mathrm{3}}$. (b) Decay of the autoionization signal at the two positions A (red triangles) and B (black dots) indicated in Fig.~\ref{fig:fig3}(b), measured at $P_{\mathrm{2}}=\mathrm{10}$~mW. The solid lines are fits using a single (red) and double (black) exponential decay. At 150~$\mu$s the signal at A has reached the noise floor.}
\end{figure}

The persistence of the narrow component, as shown in Fig.~\ref{fig:fig3}, is also consistent with the population of higher angular momentum states that have less overlap with the core, and hence a longer lifetime. Figure~\ref{fig:fig4}(a) shows the variation in the lifetime of the autoionization signal across the spectrum. The broad $\mathrm{5}s\mathrm{56}d$ component exhibits a constant lifetime, whereas the lifetime is clearly greater at the position of the narrow peak. In Fig.~\ref{fig:fig4}(b), we compare the decay of the autoionization signal at the two detunings marked in Fig.~\ref{fig:fig3}(b). At point A, away from the narrow feature, the decay is well described by a single exponential, yielding a lifetime of $\mathrm{25.0}\,\pm\,\mathrm{0.5}\ \mu$s for the \ryd state. For $\Delta t\,>\,\mathrm{50}\ \mu$s there is a clear difference between the two curves, and the decay at B deviates from a simple single exponential. The double exponential fit shown in Fig.~\ref{fig:fig4}(b) yields lifetimes of $\mathrm{24}\,\pm\,\mathrm{4}\ \mu$s and $\mathrm{60}\,\pm\,\mathrm{7}\ \mu$s, consistent with the formation of a mixture of the \ryd state and another longer-lived state. 

The nearest dipole coupled states, \pstate and $\mathrm{5}s\mathrm{54}f\,^{\mathrm{1}}F_{\mathrm{3}}$, are -15~GHz and +12~GHz from the \ryd state respectively. We can excite these states directly by applying a small electric field during the Rydberg excitation, which is switched off before the autoionization pulse. The autoionization spectra of these states are shown in Fig.~\ref{fig:fig3}(d). We measured the lifetimes of these states and found $\mathrm{64} \pm \mathrm{4}\ \mu$s for the \fstate state, and $\mathrm{84} \pm \mathrm{2}\ \mu$s for the \pstate state. The \fstate state lifetime is consistent with the second component in Fig.~\ref{fig:fig4}(b), but we observe no long-lived component at the position of the \pstate state.  

From the data presented in Figs.~\ref{fig:fig2}-\ref{fig:fig4}, we conclude that the population is transferred predominantly from the \ryd state to the \fstate state. In   Figs.~\ref{fig:fig3}(a-c), we fit the evolution of the spectrum with a two-component model. The \ryd component is described by six-channel MQDT, as shown in Fig.~\ref{fig:fig2}. The \fstate state is described by two-channel MQDT, which we fit to the data shown in  Fig.~\ref{fig:fig3}(d) to obtain the MQDT parameters. The same two-component model is fit to the spectra (Figs.~\ref{fig:fig3}(a-c)) for different values of $\Delta t$, with each component allowed to decay with the lifetimes measured in Fig.~\ref{fig:fig4}. The remaining fit parameters are the ion signals $S_D$ and $S_F$ for each component at $\Delta t=\mathrm{0.5}\ \mu$s. This two-component model describes the full time evolution of the spectrum shown in Figs.~\ref{fig:fig3}(a-c) well.

The initial ratio of population in each Rydberg state $N_F/N_D$ can be obtained from  the ratio of the ion signals $S_F/S_D$ by $N_F/N_D = (\sigma_D S_F)/(\sigma_F S_D)$, where $\sigma_{D,F}$ are the respective autoionization cross sections. The ratio of the cross-sections can be calculated from the MQDT parameters for each component \cite{MQDT}. From this analysis, we find that at $\Delta t=\mathrm{0.5}\ \mu$s and $P_{\mathrm{2}}=\mathrm{10}$~mW  (Fig.~\ref{fig:fig3}(a)), $13\,\pm\,3$\% of the Rydberg population has been transferred to the \fstate state.
 
With this quantitative analysis and the data presented in Figs.~\ref{fig:fig2}-\ref{fig:fig4}, the population transfer mechanism can be determined. We eliminate black-body radiation due to the density dependence and the rapidity of the transfer, which from Fig.~\ref{fig:fig4} appears to be complete by the end of the Rydberg excitation pulse. Similarly, Fig.~\ref{fig:fig4}(b) shows no evidence of superradiant decay. 
 
Density-dependent transfer could also result from Stark mixing due to the background ions, or dipole-dipole interactions. To estimate the importance of these two effects we have calculated the relevant dipole matrix elements using an independent electron model that neglects coupling between the singlet and triplet states. To check the validity of this method we have calculated the Stark map \cite{chan01} at $n=\mathrm{56}$ and $n=\mathrm{80}$ and found excellent agreement with experimental data. To calculate the amount of Stark mixing, the density of Rydberg atoms is also required, which we estimate by measuring the loss of ground state atoms after the Rydberg excitation. At $P_{\mathrm{2}}=\mathrm{10}$~mW, $\sim\mathrm{10}$\% of the atoms are excited to the Rydberg state.

The dominant dipole-dipole coupling is 
\begin{equation*}
\mathrm{5}s\mathrm{56}d\,^\mathrm{1}D_\mathrm{2} + \mathrm{5}s\mathrm{56}d\,^\mathrm{1}D_\mathrm{2}  \rightarrow \mathrm{5}s\mathrm{56}p\,^{\mathrm{1}}P_{\mathrm{1}} + \mathrm{5}s\mathrm{54}f\,^{\mathrm{1}}F_{\mathrm{3}}
\end{equation*}

which is 2.5~GHz from resonance. This process results in a repulsive van der Waals type interaction, and we are far from the resonant energy transfer regime. At our peak Rydberg density we estimate that the dipole-dipole interaction is responsible for $<\mathrm{0.5}$\% of state mixing. This mechanism would also populate the \pstate state, which we do not observe.

The Stark map can be used to estimate the mixing due to the field of the background ions. At our peak density, the average electric field seen by a Rydberg atom if its nearest neighbor is ionized is 0.7~V\,cm$^{-\mathrm{1}}$, which would mix in 4\% of the \fstate state and 2\% of the \pstate state. However, from the ratio of the autoionization signal to the background signal, only  1\% of the Rydberg atoms are spontaneously ionized. Therefore the field experienced by the majority of the Rydberg atoms is much lower, and the effect of Stark mixing due to the ions is negligble.

\begin{figure}
\includegraphics{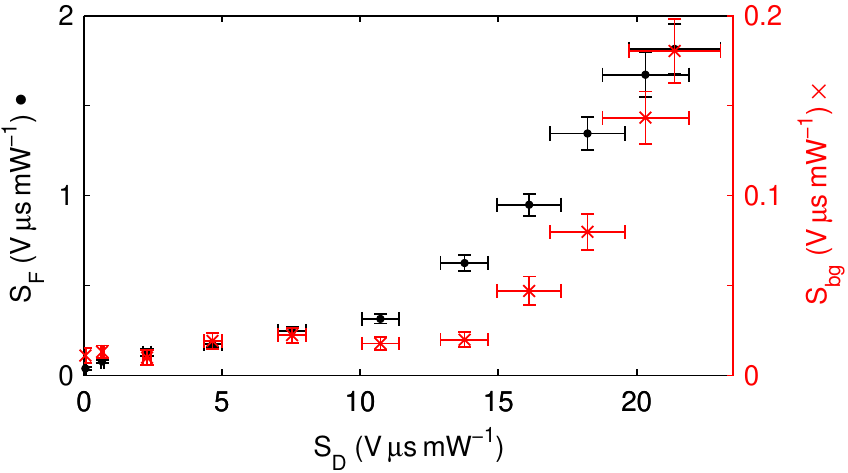}
\caption{\label{fig:fig5} (Color online) Variation in the $\Delta t = \mathrm{100}\ \mu$s autoionization ($S_F$, black dots) and background ($S_{bg}$, red crosses) signals, at position B (Fig.~\ref{fig:fig3}(b)), with $S_D$, the $\mathrm{5}s\mathrm{56}d$ autoionization signal at $\Delta t = \mathrm{0.5}\ \mu$s.}
\end{figure}

Finally, we consider the role of $l$-changing electron-Rydberg collisions. Appreciable transfer due to this mechanism must be associated with the formation of an ultra-cold plasma \cite{flannigan04}. Using autoionization as a probe, we can directly examine the link between $l$-mixing and the formation of a plasma. At a delay of $\Delta t = \mathrm{100}\ \mu$s, the autoionization signal at point B (Fig.~\ref{fig:fig3}) is dominated by the long-lived \fstate atoms. Figure~\ref{fig:fig5} shows the variation of this signal $S_F$ with the initial \ryd Rydberg signal $S_D$. The data shows a threshold in the amount of population transfer, which is accompanied by a similar threshold in the background ion signal $S_{bg}$. The rapid, spontaneous ionization of a small fraction of the atoms in Rydberg gases with repulsive interactions has been observed previously \cite{amthor07}. Once the density of the background ions is high enough, a cold plasma can form, which is accompanied by a characteristic threshold in the amount of ionization \cite{robinson00}. Under our conditions , $\sim\mathrm{1000}$ ions would be required for plasma formation \cite{killian01}, which is compatible with our estimate that 1\% of the Rydberg atoms are ionized. We conclude therefore that $l$-changing collisions due to the formation of a plasma are the dominant population transfer mechanism in these experiments.

Previously, plasma formation was observed to populate very long-lived states with $l>\mathrm{3}$. We have observed spontaneous ionization of our Rydberg gas on timescales up to 20~ms, but only at higher Rydberg densities than those discussed in this paper. Here we are probing the regime close to the threshold for plasma formation, where the ionization fraction is only $\sim$1\%. Under these conditions, we find that electron-Rydberg collisions predominantly populate the \fstate state. Higher $l$ states would give rise to even narrower, longer-lived features in the autoionization spectrum, which we do not observe.

In conclusion, an additional valence electron provides new ways to study cold Rydberg gases. The sensitivity of the autoionization spectrum to changes in angular momentum $l$ was used to observe $l$-changing electron-Rydberg collisions close to the threshold for plasma formation. Autoionization could be used to probe very short timescales, limited only by the width of the autoionizing resonance. By focusing the autoionization laser, this technique could be extended to provide simultaneous spectral, temporal and spatial resolution, which could be used to study, for example, correlations in the blockaded regime.

We thank G.~Corbett, R.~M.~Potvliege, and J.~D.~Pritchard for assistance with the Stark map calculations, and  C.~S.~Adams and D.~Carty for the loan of equipment. We also acknowledge helpful discussions with A.~Browaeys, D.~Comparat and P.~Pillet. This work was supported by EPSRC grants EP/D070287/1 and LP/82000, and Durham University.


\begin{thebibliography}{99}
\bibitem{MQDT} So that the parameters can be directly compared, we fit the \ryd and \fstate data with the same two-channel model.
\end{thebibliography}
\end{document}